\documentclass{CSML}

\def\dOi{11(4:11)2015}
\lmcsheading%
{\dOi}
{1--20}
{}
{}
{Mar.~\phantom03, 2014}
{Dec..~14, 2015}
{}

\ACMCCS{[{\bf Theory of computation}]: Logic---Finite model theory;
  [{\bf Mathematics of computing}]: Discrete mathematics---Graph theory}

\pdfoutput=1

\usepackage[czech,english]{babel}

\usepackage{hyperref}
\usepackage{xspace}

\usepackage{tikz}
\usetikzlibrary{shapes,shapes.multipart}
\tikzstyle{every picture} = [>=latex]
\usetikzlibrary{external}
\def\ca#1{{\mathcal#1}}
\let\sem\setminus
\def\prebox#1{\mathop{\mbox{#1}}}

\newcommand{\FO}{\ensuremath{\operatorname{FO}}\xspace}
\newcommand{\MSO}{\ensuremath{\operatorname{MSO}}\xspace}
\newcommand{\MSOi}{\ensuremath{\operatorname{MSO_1}}\xspace}
\newcommand{\MSOii}{\ensuremath{\operatorname{MSO_2}}\xspace}

\newcommand{\cI}{\mathcal{I}}
\newcommand{\cA}{\mathcal{A}}
\newcommand{\CC}{\mathcal{C}}
\newcommand{\II}{\mathcal{I}}
\newcommand{\TT}{\mathcal{T}}

\theoremstyle{plain}\newtheorem{claim}[thm]{Claim}

\begin{document}

\title[FO Model Checking of Interval Graphs]{FO Model Checking of Interval Graphs\rsuper*}

\author[R.~Ganian]{Robert Ganian\rsuper a}
\address{{\lsuper a}Algorithms and Complexity Group, TU Wien, Favoritenstrasse 9-11, A-1040 Vienna, Austria}
\email{rganian@gmail.com}

\author[P.~Hlin\v{e}n\'{y}]{Petr Hlin\v{e}n\'{y}\rsuper b}
\address{{\lsuper{b,d}}Faculty of Informatics, Masaryk University, Botanick\'a 68a, 
  62100 Brno, Czech Republic}
\email{\{hlineny,obdrzalek\}@fi.muni.cz}

\author[D.~Kr\'a\v{l}]{Daniel Kr\'a\v{l}\rsuper c}
\address{{\lsuper c}Mathematics Institute, University of Warwick, Coventry CV4 7AL, United Kingdom}
\email{D.Kral@warwick.ac.uk}

\author[J.~Obdr\v{z}\'{a}lek]{Jan Obdr\v{z}\'{a}lek\rsuper d}
\address{\vspace{-18 pt}}

\author[J.~Schwartz]{Jarett Schwartz\rsuper e}
\address{{\lsuper e}Computer Science Division, UC Berkeley, 387 Soda Hall
Berkeley, CA 94720-1776, United States}
\email{jarett@cs.berkeley.edu}

\author[J.~Teska]{Jakub Teska\rsuper f}
\address{{\lsuper f}Faculty of Applied Sciences, University of West Bohemia,
Univerzitn\'{\i} 8, 30614 Pilsen, Czech Republic}
\email{teska@kma.zcu.cz}


\thanks{\lsuper{a,b,c,d,f}All the authors except for Jarett Schwartz
  acknowledge support of the Czech Science Foundation under grant
  P202/11/0196.}  

\thanks{{\lsuper a}Robert Ganian acknowledges support of the FWF
  Austrian Science Fund (X-TRACT, P26696)}

\thanks{{\lsuper e}Jarett Schwartz acknowledges support of the
  Fulbright and NSF Fellowships.}

\thanks{{\lsuper c}The work of Daniel Kr\'a\v{l} on the journal
  version of this paper was also supported by the European Research
  Council under the European Union's Seventh Framework Programme
  (FP7/2007-2013)/ERC grant agreement no.~259385.}

\thanks{{\lsuper{b,d}}The work of Petr Hlin\v{e}n\'y and Jan
  Obdr\v{z}\'alek on the journal version of this paper was also
  supported by the Czech Science Foundation under grant 14-03501S}

\titlecomment{{\lsuper*}An extended abstract of an early version of this paper
  has appeared at ICALP'13.}

\begin{abstract}
We study the computational complexity of the \FO model checking problem on
interval graphs, i.e., intersection graphs of intervals on the real line.
The main positive result is that \FO model checking and
successor-invariant \FO model checking can be solved
in time $O(n\log n)$ for $n$-vertex interval graphs with representations
containing only intervals with lengths from a prescribed finite set.
We complement this result by showing that the same is not true
if the lengths are restricted to any set that is dense in an open subset,
e.g.~in the set $(1,1+\varepsilon)$.
\end{abstract}

\maketitle\hfill

\section{Introduction}

Results on the existence of an efficient algorithm for classes of
problems have recently attracted a significant amount of attention. Such results
are now referred to as algorithmic meta-theorems, also see a recent survey~\cite{kre09}.
The most prominent example is a theorem of Courcelle~\cite{cou90}
asserting that every \MSO (monadic second order) property can be model checked in linear time on the class of
graphs with bounded tree-width. Another example is a theorem of Courcelle, Makowski and
Rotics~\cite{cmr00} asserting that the same conclusion holds for graphs
with bounded clique-width when quantification is restricted to vertices and
their subsets.

In this paper, we focus on a more restricted class of graph properties,
specifically the properties expressible
in first order logic. Clearly, every such property can be tested in polynomial time
if we allow the degree of the polynomial to depend on the property of interest.
But is testing these properties {\em fixed parameter tractable}
(FPT~\cite{df13}), i.e.~are they testable in polynomial time where the degree of the polynomial does
not depend on the considered property?
The first result
in this direction could be that of Seese~\cite{see96}: every \FO property
can be tested in linear time on graphs with bounded maximum degree. A breakthrough
result of Frick and Grohe~\cite{fg01} asserts that every \FO property can be tested in almost
linear time on classes of graphs with locally bounded tree-width. Here, an almost
linear algorithm stands for an algorithm running in time $O(n^{1+\varepsilon})$
for every $\varepsilon>0$.
A generalization to graph classes locally excluding a minor (with worse running time)
was later obtained by Dawar, Grohe and Kreutzer~\cite{dgk07}. 

These results have been subsequently extended to (more general) sparse graph classes
introduced by Ne\v{s}et\v{r}il and Ossona de~Mend\'ez~\cite{no08i,no08ii,no08iii}. First Dawar and Kreutzer \cite{dk09}
(also see \cite{gk11} for the complete proof) and, independently,
Dvo\v{r}\'ak, Kr\'al' and Thomas~\cite{dkt10}, showed that every
FO property can be tested in almost linear time on classes of
graphs with locally bounded expansion; examples of such graph classes
include classes of graphs with bounded maximum degree or proper minor-closed
classes of graphs. This series of results ultimately culminated with the recent result of Grohe, Kreutzer and
Siebertz~\cite{GKS14}, who established the fixed parameter tractability of
testing \FO properties on nowhere-dense classes of graphs (nowhere-dense
being the most general class of sparse graphs).

In this work, we investigate whether structural properties of graphs that are not necessarily sparse could lead to similar results.
Specifically, we study the intersection graphs of intervals on the real line,
which are also called interval graphs. 
When restricted to unit interval graphs,
i.e.~intersection graphs of intervals with unit lengths,
one can easily deduce the existence of a linear time algorithm
for testing \FO properties from Gaifman's theorem, using the result of Courcelle et al.~\cite{cmr00} and
that of Lozin~\cite{loz08} asserting that every proper hereditary subclass of
unit interval graphs, in particular, the class of unit interval graphs
with bounded radius, has bounded clique-width.
This observation is a starting point for our research presented
in this paper.

Let us now give a definition. For a set $L$ of reals,
an interval graph is called an {\em$L$-interval graph} if it is an intersection
graph of intervals with lengths from $L$. For example, unit interval graphs
are $\{1\}$-interval graphs.
If $L$ is a finite set of rationals, then any $L$-interval graph
with bounded radius has bounded clique-width (see Section~\ref{sec:clique} for further details).
So, testing \FO properties of such graphs is fixed parameter tractable.
However, if $L$ is not a set of rationals, there exist $L$-interval graphs
with bounded radius and unbounded clique-width, 
and so the easy argument above does not apply.

Our main algorithmic result (Theorem~\ref{thm-L-interval}) says that
every fixed \FO property can be tested
in time $O(n\log n)$ for $n$-vertex $L$-interval graphs when 
$L$ is any fixed finite set of reals and an $L$-interval representation is
given on the input.
To prove this result, we employ a well-known characterization of \FO properties
by Ehrenfeucht-Fra{\"\i}ss\'e games. Specifically,
we show, using the notion of game trees introduced later, that
there exists an algorithm transforming an input $L$-interval graph to another $L$-interval
graph that has bounded maximum degree and that satisfies the same properties expressible
by \FO sentences with bounded quantifier rank.
Inspired by Engelmann, Kreutzer and Siebertz~\cite{EKS12} (also see~\cite{EKK13}),
we then extend our main algorithmic result to successor-invariant
\FO properties.
We should also mention that a recent result of Gajarsk\'y et 
al.~\cite{gajetal15} (proven subsequently after this work),
giving a fixed parameter algorithm for testing
FO properties of partial orders with bounded width, implies
Theorem~\ref{thm-L-interval} with a~running time quadratic in $n$.

On the negative side, we show that if $L$ is an (infinite) set that is dense
in some open set, then $L$-interval graphs can be used to model arbitrary graphs.
Specifically, we show that $L$-interval graphs for these sets $L$ allow
efficient polynomially bounded \FO interpretations of all graphs. Consequently,
testing \FO properties for $L$-intervals graphs for such sets $L$ is 
W[2]-hard (see Corollary~\ref{cor-FO}) and hence unlikely to be fixed parameter tractable.
In addition, we show that unit interval graphs allow an efficient polynomially bounded
MSO interpretation of all graphs and a successor \FO interpretation of all graphs.
So, our main algorithmic result cannot be extended to any of these two stronger logics.

The paper is organized as follows.
In Section~\ref{sec:preliminary}, we introduce the notation and the computational
model used in the paper. In the next section, we present an $O(n\log n)$
algorithm for deciding \FO properties of $L$-interval graphs for finite sets $L$,
and we extend this result to successor-invariant \FO properties
in Section~\ref{sec:succinvar}.
Then, we present proofs of the facts mentioned above
on the clique-width of $L$-interval graphs with bounded radius in Section~\ref{sec:clique}.
We finish with the several results on the interpretability of all graphs in interval graphs in Section~\ref{sec:interpret}.

\section{Preliminaries}
\label{sec:preliminary}

An {\em interval graph} is a graph $G$ such that
every vertex $v$ of $G$ can be associated with an interval $J(v)=[\ell(v),r(v))$ such that
two vertices $v$ and $v'$ of $G$ are adjacent if and only if $J(v)$ and $J(v')$ intersect
(it can be easily shown that the considered class of graphs remains the same regardless of whether we consider open,
half-open or closed intervals in the definition).
We refer to such an assignment of intervals to the vertices of $G$ as a {\em representation} of $G$.
The point $\ell(v)$ is the {\em left end point} of the interval $J(v)$ and $r(v)$ is its {\em right end point}.

If $L$ is a set of reals and $r(v)-\ell(v)\in L$ for every vertex $v$,
we say that $G$ is an {\em $L$-interval} graph and
we say that the representation is an {\em $L$-representation} of $G$.
For example, if $L=\{1\}$, we speak about unit interval graphs.
Finally, if $r(v)-\ell(v)\in L$ and $0\le\ell(v)\le r(v)\le d$ for some real $d$,
i.e.~all intervals are subintervals of $[0,d)$, we speak about {\em $(L,d)$-interval graphs}.
Note that if $G$ is an interval graph of radius $k$, then $G$ is also an $(L,(2k+1)\max L)$-interval graph (we
use $\max L$ to denote the maximum element of the set $L$).

While an (unrestricted) interval representation of a given interval graph
$G$ can be found in linear time~\cite{bl76} and the
same applies to unit interval graphs~\cite{cknos95},
there seem to be no results in the literature about the complexity of
finding an $L$-representation of a given $L$-interval graph when 
$L$ is a finite set of positive reals and~$|L|>1$.
Although, Pe'er et al.~\cite{ps97} prove that a related
interval graph recognition problem in that every vertex of the input graph 
comes together with its prescribed interval length is NP-hard.
We thus suspect that the recognition problem of $L$-interval graphs might be
hard in the computational complexity sense as well and, consequently, 
we always assume in this paper that an
input graph comes alongside with its $L$-representation.

We now introduce two technical definitions related to manipulating intervals and their lengths.
These definitions are needed in the next section.
If $L$ is a set of reals, then $L^{(k)}$ is the set of all integer linear combinations of numbers from $L$
with the sum of the absolute values of their coefficients bounded by $k$. For instance, $L^{(0)}=\{0\}$ and
$L^{(1)}=L\cup(-L)\cup\{0\}$. An {\em $L$-distance} of two intervals $[a,b)$ and $[c,d)$ is the smallest $k$
such that $c-a\in L^{(k)}$. If no such $k$ exists, then the $L$-distance of two intervals
is defined to be $\infty$.

Since we do not restrict our attention to $L$-interval graphs where $L$ is a set of rationals,
we should specify the computational model considered. We use the standard RAM model
with infinite arithmetic precision and unit cost of all arithmetic operations.
However, we refrain from trying to exploit the power of this computational model
by encoding other data in the infinite precision variables to manipulate the time complexity of the presented algorithms.
In particular,
we only store the end points of the intervals of the representations of input graphs and their differences
in numerical variables with infinite precision and compare these values, e.g.~to decide the vertex adjacencies.

\subsection{Parameterized Complexity}\label{section:pc}

Next we give a very brief review of the most important
concepts of parameterized complexity. For an in-depth treatment of the
subject we refer the reader to other sources, e.g.~\cite{df13}.

The instances of a parameterized problem can be considered as pairs
$\langle I,k\rangle$ where $I$ is the \emph{main part} of the instance and $k$ is
the \emph{parameter} of the instance; the latter is usually a
non-negative integer.  A parameterized problem is
\emph{fixed parameter tractable (FPT)} if instances $\langle I,k\rangle$ 
of size $n$ (with respect to some reasonable encoding) can be solved
in time $O(f(k)\cdot n^c)$ where $f$ is a computable function and $c$
is a constant independent of $k$.
In the area of {\em parameterized model checking}, instances are
considered in the form $\langle(G,\phi),|\phi|\rangle$
where $G$ is a structure, $\phi$ a formula, the question is whether
$G\models\phi$ and the parameter is the size of~$\phi$.
Therefore, when speaking about parameterized complexity
of \FO model checking we implicitly consider the formula size as a parameter.

The framework of parameterized complexity offers a
completeness theory, similar to the theory of NP-completeness, 
that allows the accumulation of strong theoretical
evidence that a parameterized problem is not fixed parameter
tractable.  This completeness theory is based on the \emph{weft
  hierarchy} of equivalence classes W[1],W[2],\dots, W[P] of
certain parameterized decision problems under \emph{parameterized
  reductions}. A parameterized reduction is an
extension of a polynomial-time many-one reduction 
to parameterized problems that ensures that the parameter of the new instance
is bounded by a function of the parameter of the original instance.
It is known that, unless the Exponential Time Hypothesis fails~\cite{ipz01},
W[1]-hard problems are not fixed parameter tractable.

The class AW[*] extends the weft hierarchy by adding the notion of
alternations, and is formally based on the problem of deciding the
satisfiability of quantified boolean formulas. 
In particular, AW[*]-hard problems are also W[1]- and W[2]-hard.
Showing that a parameterized problem is AW[*]-hard hence provides
a very solid evidence that the problem is not fixed parameter tractable.
The parameterized \FO model checking problem on general structures 
as well as on all graphs is AW[*]-complete~\cite{dft96}.

There exists an even stronger notion of hardness for parameterized
problems: a parameterized problem is \emph{para-\textup{NP}-hard}
if there exists a parameter $k_0$ such that the problem restricted to
the instances $\langle I,k_0\rangle$ of parameter value equal to~$k_0$ is NP-hard.

\subsection{Clique-width}

We now briefly present the notion of clique-width, introduced in~\cite{co00}.
A $k$-labeled graph is a graph whose vertices are assigned integers (called
labels) from $1$ to $k$ (each vertex has precisely one label).  The {\em
  clique-width} of a graph $G$ equals the minimum $k$ such that $G$ can be
obtained using the following four operations: creating a vertex labeled 
$1$, relabeling all vertices with label $i$ to label $j$, adding all edges between
the vertices with label $i$ and the vertices with label $j$, and
taking a disjoint union of graphs obtained using these operations.

\subsection{First Order Properties}

In this subsection,
we introduce concepts from logic and model theory which we use. A {\em first order (FO) sentence}
is a formula with no free variables with the usual logical connectives and quantification allowed only
over variables for elements (vertices in the case of graphs).
A {\em monadic second order (MSO) sentence}
is a formula with no free variables with the usual logical connectives where, unlike in \FO sentences, quantification over subsets
of elements is allowed. An {\em \FO property} is a property expressible by an \FO sentence; similarly,
an {\em \MSO property} is a property expressible by an \MSO sentence.
Finally, the {\em quantifier rank} of a formula is the maximum number of nested quantifiers.

\FO sentences are closely related to the so-called Ehrenfeucht-Fra{\"\i}ss\'e games.
The {\em $d$-round Ehrenfeucht-Fra{\"\i}ss\'e game} is played on two relational structures $R$ and $R'$ (of the same type)
by two players, referred to as the {\em spoiler} and the {\em duplicator}. 
In each round $i=1,2,\dots d$,
the spoiler chooses an element in one of the structures and the duplicator chooses an element in the other.
Let $x_i$ and $y_i$ be the elements of $R$ and $R'$ chosen in the $i$-th round.
We say that the duplicator {\em wins} the game if there is a strategy for the duplicator such that, for any strategy of the spoiler, 
the substructure of $R$ induced by the elements $x_1,\ldots,x_d$ is always isomorphic
to the substructure of $R'$ induced by the elements $y_1,\ldots,y_d$, with the isomorphism mapping each $x_i$ to $y_i$.

The following theorem~\cite{ehr61,fra54} relates Ehrenfeucht-Fra{\"\i}ss\'e games to \FO sentences of quantifier rank at most $d$.

\begin{thm}
\label{thm-EF}
Let $d$ be an integer. The following statements are equivalent for any two structures $R$ and $R'$:
\begin{itemize}
\item The structures $R$ and $R'$ satisfy the same \FO sentences of quantifier rank at most $d$.
\item The duplicator wins the $d$-round Ehrenfeucht-Fra{\"\i}ss\'e game for $R$ and $R'$.
\end{itemize}
\end{thm}

We describe possible courses of the $d$-round Ehrenfeucht-Fra{\"\i}ss\'e games by rooted trees.
A {\em $d$-EF-tree $\TT$} is a rooted tree with the following properties:
\begin{enumerate}
\item each leaf $v$ of $\TT$ is associated with a relational structure $S(v)$ with elements labelled with $1,\ldots,d$ such that each element of $S(v)$ has at least one label (but possibly more labels) and each label is used exactly once, and
\item all the leaves of $\TT$ are at depth $d$.
\end{enumerate}
The {\em full $d$-EF-tree $\TT_R$} of a relational structure $R$ is a $d$-EF-tree $\TT$ such that
\begin{enumerate}
\item the edges from each internal node $u$ to its descendants are in one-to-one correspondence with the elements of $R$, and
\item the structure $S(v)$ associated with a leaf $v$ of $\TT_R$ is the substructure of $R$ induced by the elements corresponding to the edges on the unique path from the root to $v$ and the element corresponding to the $i$-th edge of this path is labelled by $i$.
\end{enumerate}
A mapping $f$ from a $d$-EF-tree $\TT$ to another $d$-EF-tree $\TT'$ is an {\em EF-homomor\-phi\-sm} if the following three conditions hold:
\begin{enumerate}
\item if $u$ is the parent of a vertex $v$ of $\TT$, then $f(u)$ is the parent of $f(v)$ in $\TT'$,
\item if $u$ is a leaf of $\TT$, then $f(u)$ is a leaf of $\TT'$, and
\item the relational structures associated with $u$ and $f(u)$ are the same.
\end{enumerate}
Two $d$-EF-trees $\TT$ and $\TT'$ are {\em EF-equivalent}
if there exist an EF-homomorphism from $\TT$ to $\TT'$ and an EF-homomorphism from $\TT'$ to $\TT$.
An EF-homomorpishm that is bijective is an EF-isomorphism.

We now formalize the connection between $d$-EF-trees and Ehrenfeucht-Fra{\"\i}ss\'e games. 

\begin{thm}
\label{thm-EF-tree}
Let $d$ be an integer and let $R$ and $R'$ be two relational structures.
If the full $d$-EF-trees of $R$ and $R'$ are EF-equivalent,
then the duplicator wins the $d$-round Ehrenfeucht-Fra{\"\i}ss\'e game for $R$ and $R'$.
\end{thm}

\proof
Let $\TT$ and $\TT'$ be the $d$-EF-trees for $R$ and $R'$, respectively, and
let $f:\TT\to\TT'$ and $f':\TT'\to\TT$ be the EF-homomorphisms witnessing their EF-equivalence. We claim
that the duplicator wins the $d$-round Ehrenfeucht-Fra{\"\i}ss\'e game, using the following strategy:
In the first round, if the spoiler chooses $x_1$ in $R$,
then the duplicator responds with $y_1=f(x_1)$.
If the spoiler chooses $y_1$ in $R'$, the duplicator responds with $x_1=f'(y_1)$.
Assume that the $i-1$ rounds of the game have been played,
the elements chosen in the structures $R$ and $R'$ are $x_1,\ldots,x_{i-1}$ and $y_1,\ldots,y_{i-1}$, respectively, and
the spoiler chooses an element $x_i$ in $R$.
Let $u_0,\ldots,u_i$ be the path in $\TT$ formed by the edges corresponding to $x_1,\ldots,x_i$.
The duplicator chooses the element $y_i$ of $R'$ that corresponds to the edge $f(u_{i-1})f(u_i)$ in $\TT'$.
The definitions of full $d$-EF-trees and an EF-homomorphism yield that
the substructures of $R$ and $R'$ induced by $x_1,\ldots,x_i$ and $y_1,\ldots,y_i$ are isomorphic
through the isomorphism mapping $x_j$ to $y_j$, $1\le j\le i$.
In particular, they are isomorphic after the $d$ rounds of the game and
the duplicator wins.
\qed

The converse implication, i.e.~that if the duplicator wins the $d$-round Ehren\-feucht-Fra{\"\i}ss\'e game for $R$ and $R'$,
then the $d$-EF-trees for the game played on relational structures $R$ and $R'$ are EF-equivalent, is also true.
However, we omit the proof since we only need the implication given by Theorem~\ref{thm-EF-tree}.
We show that full $d$-EF-trees can pruned to be of bounded size.

\begin{lem}
\label{lem:prune}
Consider a fixed type of relational structures.
Every class of EF-equivalent $d$-EF-trees contains a unique tree (up to an EF-isomorphism) with the minimum number of leaves and
the number of non-EF-equivalent $d$-EF-trees is finite.
\end{lem}

\proof
Let $\TT$ and $\TT'$ be EF-equivalent $d$-EF-trees with the minimum number of leaves.
Suppose that there exists a non-bijective EF-homomorphism $f$ from $\TT$ to $\TT'$.
Let $f'$ be an EF-homomorphism from $\TT'$ to $\TT$.
Let $\TT''$ be the $d$-EF-tree that is the subtree of $\TT'$ induced by the image of $f$.
Since $f$ is an EF-homomorphism from $\TT$ to $\TT'$ and
$f'$ restricted to the image of $f$ is an EF-homomorphism from $\TT''$ to $\TT$,
the $d$-EF-tree $\TT''$ is a $d$-EF-tree EF-equivalent to $\TT$ with the smaller number of leaves.

To show that the number of non-EF-equivalent $d$-EF-trees is finite,
we describe the minimal elements of EF-equivalence classes in a constructive way.
Let $\TT$ be a $d$-EF-tree.
If a vertex of $\TT$ at depth $d-1$ is adjacent to two leaves associated with the same labelled structure,
delete one of them.
The original $d$-EF-tree has a $d$-EF-homomorphism to the new one: map all the vertices except the deleted one to themselves and
map the deleted leaf to the other leaf associated with the same labelled structure.
After this operation, the number of children of any vertex at depth $d-1$ does not exceed the number of non-isomorphic
structures with their vertices labelled by $1,\ldots,d$; let $K$ be this number.
Now, if any vertex has two children such that their subtrees are isomorphic (preserving the labelled structures associated with their leaves),
deleting one of them with its subtree results in a $d$-EF-tree EF-equivalent to $\TT$.
When the pruning process stops, we have obtained the minimal $d$-EF-tree EF-equivalent to $\TT$ (a non-injective EF-homomorphism
from a $d$-EF-tree always exhibits a vertex that can be pruned in the described way).

After pruning $\TT$ in the way we described, every vertex at depth $d-2$ has at most $2^K$ children,
every vertex at depth $d-3$ has at most $2^{2^K}$ children, etc.
So, every EF-equivalence class contains a $d$-EF-tree of size bounded by a function of $K$ and $d$.
Clearly, there can be only finitely many such such $d$-EF-trees.
\qed

In what follows, we will refer to the minimal $d$-EF-tree EF-equivalent to the full $d$-EF-tree of a relational structure $R$ as
the {\em$d$-EF-tree of a relational structure $R$}.
Note that the $d$-EF-tree of a relational structure $R$ can be constructed from the full $d$-EF-tree in an efficient way
through the pruning process described in the proof of Lemma~\ref{lem:prune}.

\section{FO Model Checking}
Using Theorems~\ref{thm-EF} and~\ref{thm-EF-tree},
we prove the following result for $L$-interval graphs.

\begin{thm}
\label{thm-kernel}
For every finite subset $L$ of reals and every integer $d\geq0$, there exist an integer $K_0$ and
an algorithm $\cA$ with the following properties. The input of $\cA$ is an $L$-representation
of an $n$-vertex $L$-interval graph $G$ and $\cA$ outputs in time $O(n\log n)$
an $L$-representation of an induced subgraph $G'$ of $G$ such that
\begin{itemize}
\item every unit interval contains at most $K_0$ left end points of the intervals
      corresponding to vertices of $G'$, and
\item $G$ and $G'$ satisfy the same \FO sentences with quantifier rank at most $d$.
\end{itemize}
\end{thm}

\proof
We are going to use Ehrenfeucht-Fra{\"\i}ss\'e games to (possibly) identify an interval
representing a vertex of~$G$ that can be deleted without changing the set
of \FO sentences of quantifier rank at most $d$ satisfied by the input graph.
Hence, we first focus on proving the existence of the number $K_0$ and the subgraph $G'$ and
we postpone the algorithmic considerations to the end of the proof.

We start with perturbing the intervals to guarantee that all the left end points of the intervals representing the vertices of $G$ are distinct.
Choose $\delta$ to be the minimum distance between distinct end points of the intervals in the representation.
Sort the intervals by their left end points (resolving ties arbitrarily) and
shift the $i$-th interval by $i\delta/2n$, for $i=1,\ldots,n$, to the right.
This does not change the graph represented by the intervals and all the end points become distinct.
Note that this pertubration can be simulated by storing each end point in the form $(x,i)$ where $x$ is its original coordinate;
the pair $(x,i)$ represents the point $x+i\delta/2n$ and the lexicographic ordering of the pairs is to the ordering of the modified end-points.
In this way, we can perform the perturbation in a way consistent with our computational model,
i.e., without actually modifying the positions of the end points.

Choose $\varepsilon$ to be the minimum positive element of $L^{(2^{d+2})}$. We now establish the following.
\begin{claim}
There exists a number $K$ depending only on $L$ and $d$ such that
if any interval $[a,a+\delta)$, $\delta\le\varepsilon$, contains more than $K$ left end points of the intervals representing the vertices of $G$,
then $G$ has a vertex $w$ such that $G$ and $G\setminus\{w\}$ satisfy the same FO sentences with quantifier rank at most $d$.
\end{claim}
Fix $[a,a+\delta)$. Let $\II$ be the set of all intervals $[x,x+\delta)$ such that $x-a\in L^{(2^{d+1})}$.
By the choice of $\varepsilon$, the intervals of $\II$ are disjoint.
In addition, the set $\II$ is finite (recall that $L$ is finite).
Let $W$ be the set of vertices $w$ of $G$ such that the left end point $\ell(w)$ of the interval corresponding to $w$ is in an interval from $\II$.
For $w\in W$, let $i(w)$ be the left end point of the interval from $\II$ containing $\ell(w)$.
Define a linear order on $W$ such that $w\le w'$ for $w\not=w'$ from $W$ if
\begin{itemize}
\item $\ell(w)-i(w)<\ell(w')-i(w')$, or
\item $\ell(w)-i(w)=\ell(w')-i(w')$ and $\ell(w)<\ell(w')$.
\end{itemize}
We view $W$ as a linearly ordered set with each of its elements colored (associated) with the pair formed by $i(w)$ and the length of the interval of $w$,
i.e. with elements of $\II\times L$.
Observe that the colors of the elements of $W$ (together with the linear order) determine the subgraph of $G$ induced by $W$.

Let $K$ be the sum of the number of edges of all non-EF-isomorphic minimal $d$-EF-trees for Ehrenfeucht-Fra{\"\i}ss\'e games
played on linearly ordered sets with elements colored with $\II\times L$.
The number $K$ is well defined (finite) by Lemma~\ref{lem:prune}.
If $W$ contains more than $K$ elements, then there is an element $w\in W$ such that
the $d$-EF-trees of $W$ and $W\setminus\{w\}$ are the same, i.e. 
the duplicator wins the $d$-round Ehrenfeucht-Fra{\"\i}ss\'e game by Theorem~\ref{thm-EF-tree}.
Fix such $w$ for the rest of the proof.

We now describe a strategy for the duplicator to win the $d$-round Ehrenfeucht-Fra{\"\i}ss\'e game
for the graphs $G$ and $G\setminus w$.
During the game, some intervals from $\II$ will be marked as {\em altered}.
At the beginning, the only altered interval is the interval $[a,a+\delta)$.

The duplicator strategy in the $i$-th round of the game is the following.
\begin{itemize}
\item If the spoiler chooses a vertex $u$ with $\ell(u)$ in an interval of $\II$ at $L$-distance
      at most $2^{d+1-i}$ from an altered interval, then the duplicator follows the winning strategy
      for the $d$-round Ehrenfeucht-Fra{\"\i}ss\'e game for the linearly ordered colored sets $W$ and $W\setminus\{w\}$.
      This gives a vertex $v$ to choose in the other graph.
      In addition, the duplicator marks the interval of $\II$ that contains $\ell(u)$ as altered (note that
      $\ell(u)$ and $\ell(v)$ necessarily belong to the same interval of $\II$).
\item Otherwise, the duplicator chooses the same vertex in the other graph and no new intervals are marked as altered.
\end{itemize}

\noindent We now argue that the subgraphs of $G$ and $G\setminus w$ obtained in this way are isomorphic.
Let $U=\{u_1,\ldots,u_d\}$ be the chosen vertices of $G$ and $U'=\{u'_1,\ldots,u'_d\}$ those chosen in $G\setminus w$.
Let us refer to the vertices corresponding to the intervals with left end points in the altered intervals as
{\em altered vertices}.
If $u_i$ is not altered, then $u_i=u'_i$.
If $u_i$ is altered, then $\ell(u_i)$ and $\ell(u'_i)$ belong to the same interval $J\in\II$.
Suppose two vertices $u_j$ and $u'_j$ are adjacent differently to $u_i$ than to $u'_i$.
Then $\ell(u_j)$ and $\ell(u'_j)$ belong to an interval $J'\in\II$ at $L$-distance at most one from $J$.
Observe that the $L$-distance of $J'$ from an altered interval in the $i'$-th round, $i'<i$, is at most $2^{d+1-i'}$.
Hence, if $j<i$, then $u_j$ and $u'_j$ are altered because $J'$ was at $L$-distance at most $2^{d+1-j}$ in the $j$-th round.
If $j>i$, then $u_j$ and $u'_j$ are altered
because the interval $J$ turned to be altered in the $i$-th round and the $L$-distance of $J$ and $J'$ is at most one.

Since we have followed a winning strategy for the duplicator for the sets $W$ and $W\setminus\{w\}$,
the colors of $u_j$ and $u'_j$ are the same and they are comparable to $u_i$ and $u'_i$ in the same way.
In particular, they are adjacent to $u_i$ and $u'_i$ in the same way.
We conclude that the duplicator wins the game, which finishes the proof of the claim.

We now show that the statement of the theorem is true with $K_0=K\lceil\varepsilon^{-1}\rceil$.
The algorithm sorts the left end points of all the intervals (this requires $O(n\log n)$ time) and
for each of these points computes the distance to the left end of the interval 
that is $K$ positions to the right in the obtained order.
If all these distances are at least $\varepsilon$,
then every interval of length at most $\varepsilon$ contains at most $K$ left end points of
the intervals and the representation is of the desired form.

Otherwise, we choose $a$ and $b$ with the smallest $b-a$ such that the interval $[a,b)$ contains $K+1$ points and $b-a<\varepsilon$.
By the choice of this interval, any interval of length $b-a$
contains at most $K+1$ left end points of the intervals from the representation.
So, the size of the $d$-EF-tree for the game played on the vertices $v$ with $\ell(v)$ in the intervals at $L$-distance at most $2^{d+1}$
from $[a,b)$ is bounded by a function of $K$, $d$ and $|L|$.
Since this quantity is independent of the input graph, we can identify (in constant time) a vertex $w$ with $\ell(w)\in [a,b)$
with the properties from the claim. We delete this vertex from the graph $G$.
We then update the order of the left end points and the at most $K$ computed distances affected by removing $w$, and
iterate the whole process.
Storing the distances in a heap results in an algorithm that needs $O(\log n)$ per vertex removal.
Hence, the running time of the algorithm is bounded by $O(n\log n)$.
\qed

It is possible to think of several strategies to efficiently decide \FO properties of $L$-interval graphs
given Theorem~\ref{thm-kernel}.
We present one of them.
Fix an \FO sentence $\Phi$ with quantifier rank $d$ and
apply the algorithm from Theorem~\ref{thm-kernel} to get an $L$-interval graph
and a representation of this graph such that every unit interval contains at most
$K_0$ left end points of the intervals of
the representation. After this preprocessing step, every vertex of the new graph
has at most $K_0\cdot\lceil\max L\rceil$ neighbors. 
In particular, the maximum degree of the new graph is bounded.
The result of Seese~\cite{see96} asserts that every \FO property can be decided in linear time for graphs with bounded maximum degree, and so we conclude:

\begin{thm}
\label{thm-L-interval}
For every finite subset $L$ of reals and every \FO sentence $\Phi$, there exists an algorithm
running in time $O(n\log n)$ that decides whether an input $n$-vertex $L$-interval graph $G$
given by its $L$-representation satisfies $\Phi$.
\end{thm}

\section{Successor-invariant FO}
\label{sec:succinvar}

A successor relation on $X$ is simply a directed path on the vertex set~$X$.
An \FO sentence over a successor-equipped relational structure is 
{\em successor-invariant} if its truth does not change when the same
structure is equipped with a different successor relation.
Successor-invariant \FO sentences are generally more expressive than \FO sentences~\cite{Ros07}.
However, our previous result can be extended to this more expressive setting.

A useful tool when solving the model checking problem on a class of
structures is the ability to ``efficiently translate'' an instance of the problem to a different class of structures.
This tool is formalized through the concept of interpretability of logic theories~\cite{Rab64}.
An {\em\FO graph interpretation} is a pair $\cI=(\nu,\mu)$ of {\em \FO} formulas $\nu$ and $\mu$ with $1$ and $2$ free variables, respectively.
If $G$ is a graph, then $\cI(G)$ is the graph such that
\begin{itemize}
\item its vertex set is the set of all $v\in V(G)$ such that $G\models \nu(v)$, and
\item its edge set is the set of all the pairs $u$ and $v$ such that $G\models \nu(u)\wedge\nu(v)\wedge\mu(u,v)$.
\end{itemize}
We require that the edge set relation as defined must be symmetric,
i.e.~$G\models (\nu(u)\land\nu(v))\Rightarrow(\mu(x,y)\Leftrightarrow\mu(y,x))$ for every graph $G$.

Similarly, an {\em \FO successor-graph interpretation} is a triple $\cI=(\nu,\mu,\sigma)$ of {\em \FO} formulas
where $\nu$, $\mu$ and $\sigma$ have one, two and two free variables, respectively.
The meaning of $\nu$ and $\mu$ is the same and $\sigma$  should represent the successor relation:
$v$ is the successor of $u$ iff $G\models\nu(u)\wedge\nu(v)\wedge\sigma(u,v)$.
Analogously, one may also define an \MSO graph interpretation
where $\nu$ and $\mu$ are allowed to be \MSO formulas.

A class $\CC_1$ of (successor-equipped) graphs has an \FO interpretation in a class $\CC_2$ of graphs
if there exists an \FO (successor-)graph interpretation $\cI$ such that every (successor-equipped) graph $G_1$ from $\CC_1$
is isomorphic to $\cI(G_2)$ for some $G_2\in\CC_2$.
An interpretation is {\em efficient} if it can be computed in polynomial time.
If $h$ is an integer function, then $\cI$ is $h$-bounded
if there exists such $G_2$ for every $G_1$ with $|V(G_2)|\le h(|V(G_1)|)$.
In particular, if $h$ is a linear function, then we say that $\cI$ is {\em linearly bounded} and
if $h$ is a polynomial function, then we say that $\cI$ is {\em polynomially bounded}.

\begin{thm}
\label{thm-L-interval-succ}
For every finite subset $L$ of reals and every successor-invariant \FO sentence $\Phi$, there exists an algorithm
running in time $O(n\log n)$ that decides whether an input $n$-vertex $L$-interval graph $G$
given by its $L$-representation satisfies $\Phi$.
\end{thm}

\proof
The straightforward criterion~\cite[Lemma~5.3]{EKS12} implies that
it is enough to construct an efficient linearly bounded \FO successor-graph
interpretation of the class of $L$-interval graphs equipped with a suitable
successor relation in the class of $L$-interval graphs and apply Theorem~\ref{thm-L-interval}.

Before proceeding further with the proof, we need two definitions.
Two vertices in a graph are {\em twins} if their neighborhoods are the same.
An interval representation is {\em nice} if each interval except the last interval contains the left end point of another interval.
Note that not all interval graphs have nice representations (e.g.~disconnected graphs do not).

As in the proof of Theorem~\ref{thm-kernel}, we first perturb the intervals so that all their end points are distinct.
First suppose that the $L$-interval representation of $G$ is nice and
let $G^+$ be the graph $G$ equipped the the successor relation given by 
the ordering of the left end points of the intervals.
Notice that if a vertex $y$ is the successor of a vertex $x$ in $G^+$, then
$x,y$ are adjacent in~$G$.
We now construct an \FO successor-graph interpretation $\cI_1$ in $L$-interval graphs with intervals colored black, red, green and blue.

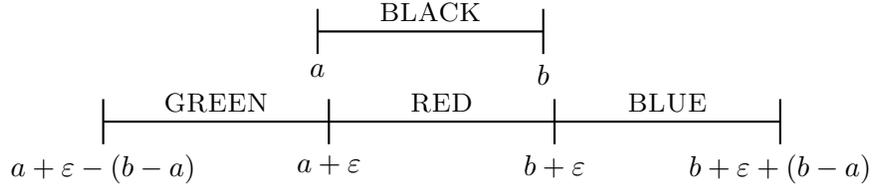
\begin{figure}[!htbp]
  \centering

\begin{tikzpicture}[scale=3]
  \draw[-, thick] (.95,.4) -- (1.95,.4);
  \foreach \x/\xtext in {.95/$a$,1.95/$b$}
    \draw[thick] (\x,.5) -- (\x,.3) node[below] {\xtext};
   
  \draw (1.45,.4) node[above] {\small\textsc{BLACK}};

  \draw[-, thick] (0,0) -- (3,0);
\foreach \x/\xtext in {0/$a+\varepsilon-(b-a)$,1/$a+\varepsilon$,2/$b+\varepsilon$, 3/$b+\varepsilon+(b-a)$}
    \draw[thick] (\x,.1) -- (\x,-0.1) node[below] {\xtext};
\foreach \x/\xtext in {0.5/\small\textsc{GREEN}, 1.5/\small\textsc{RED}, 2.5/\small\textsc{BLUE}}
  \draw (\x,0) node[above] {\xtext};
\end{tikzpicture}
  
  \caption{The intervals representing the four vertices of $G_1$ corresponding to a vertex.}
  \label{fig:colorful}
\end{figure}

Fix $G^+$ and let us start with constructing the colored $L$-interval graph, which we call $H$.
Let $\varepsilon>0$ be such that any two end points of the intervals
in the representation of $G$ are at distance larger than $3\varepsilon$.
For each interval $[a,b)$, the interval representation of $H$ contains the following four intervals (see Figure~\ref{fig:colorful}):
\begin{itemize}
\item the black interval $[a,b)$,
\item the green interval $[2a-b+\varepsilon,a+\varepsilon)$,
\item the red interval $[a+\varepsilon,b+\varepsilon)$, and
\item the blue interval $[b+\varepsilon,2b-a+\varepsilon)$.
\end{itemize}
If $v$ is the vertex of $G^+$ corresponding to $[a,b)$, the four vertices corresponding to the intervals above
are denoted by $v_K$, $v_G$, $v_R$ and $v_B$, respectively. Observe that $H$ has no twins.

We now define the interpretation $\cI_1=(\nu_1,\mu_1,\sigma_1)$.
The relations $\nu_1$ and $\mu_1$ are defined as
\begin{align}\label{eq:numupred}
\nu_1(x)\equiv \prebox{black}(x) \mbox{~~and~~}
	\mu_1(x,y)\equiv \prebox{edge}(x,y)
.\end{align}
The definition of $\sigma_1$ is more involved.
For a vertex $v_K\in V(H)$, the red vertex $v_R$ has the same neighborhood as $v_K$ except for the green vertex $v_G$.
Note that $v_R$ is the only red vertex adjacent to $v_K$ with this property: indeed,
any other red vertex $u_R$ adjacent to $v_K$ is distinguished from $v_R$ by the adjacency to $v_B$ or $u_B$.
Hence, every black vertex $v_K$ can be uniquely associated with the green vertex $v_G$ by an \FO formula $\prebox{assoc}(x,y)$.
In particular, $\prebox{assoc}(x,y)$ holds only if $x=v_K$ and $y=v_G$.

If the intervals of the black vertices $u_K$ and $v_K$ intersect,
then the inequality $\ell(u_K)<\ell(v_K)$ can be captured by an \FO formula $\prebox{less}(u_K,v_K)$.
Specifically, this inequality can be expressed as
\begin{align}\label{eq:lesspred}
\prebox{less}(x,y)\equiv\, x\not=y\wedge\prebox{edge}(x,y)\wedge
	\exists z\big[\prebox{assoc}(x,z)\wedge\neg \prebox{edge}(z,y)\big]
.\end{align}
The successor relation can now be interpreted using \eqref{eq:lesspred} as follows.
\begin{align}\label{eq:succ1}
\sigma_1(x,y)\equiv\, \prebox{less}(x,y) \wedge
	\forall z\big[\neg\prebox{black}(z)\vee 
		\neg\prebox{less}(x,z)\vee\neg\prebox{less}(z,y)\big]
\end{align}

We now adapt the construction to the case when the $L$-representation of $G$ is not nice.
To do so, we introduce a fifth color, which we will refer to as gray.
If there is an interval $J$ that is not the last interval and that does not contain the left end point of another interval,
we insert a gray interval $J'$ of length $\max L$ that has its left end point inside $J$.
If $J'$ does not contain the end point of another interval,
we can shift all the intervals to the right from $J'$ by the same distance in such a way that
the left end point of one of them, say $J''$, moves inside $J'$ and
the only new intersection we have introduced is the one between $J'$ and $J''$.

After this modification, we perform the construction described earlier,
replacing each original interval with black, green, red and blue intervals and
each gray interval with gray (in the role of the black interval), green, red and blue intervals.
Let $H$ be the graph obtained in this way.
The number of black intervals in the representation of $H$ is the number of vertices of $G$.
Since there is the left end point of a black interval between the left end points of any two gray intervals,
the number of gray intervals is at most the number of black intervals.
Finally, the numbers of green, red and blue intervals are the same and
they are equal to the total number of black and gray intervals.
We conclude that $H$ has at most $8|V(G)|$ vertices.

It remains to adapt the \FO successor-graph interpretation $\cI_1$,
in particular, the \FO formula $\sigma_1$.
The successor relation between the black intervals is again given by the order of their left end points.
Since there is the left end point of at most a single gray interval between any two consecutive left end points of black intervals,
we can define the interpretation of the successor relation as follows:
$$\sigma_1'(x,y)\equiv\sigma_1(x,y)\vee \exists z\big[
 \prebox{gray}(z)\wedge\sigma_1(x,z)\wedge\sigma_1(z,y)\big]
.$$
Observe that $H$ has no twins.

We now construct a \FO graph interpretation $\cI_2$ of five-colored $L$-interval graphs
with no twins in $L$-interval graphs.
Every gray, green, red and blue interval is replaced with two, three, four or five
identical uncolored copies; black intervals only lose their color.
Let $H'$ be the constructed $L$-interval graph.
Observe that the number of vertices of $H'$ is at most $27|V(G)|$.

Since $H$ has no twins, the vertices of $H'$ corresponding to the black intervals can be identified
by
$\prebox{black}(x)\equiv \forall y\big[
 x=y\vee \exists z \prebox{edge}(x,z)\not\Leftrightarrow\prebox{edge}(y,z)
\big]$.
In a similar way, one may define \FO formulas $\prebox{gray}(x)$, $\prebox{green}(x)$, $\prebox{red}(x)$ and $\prebox{blue}(x)$
to express that the vertex $x$ is one of the twins (of multiplicity two, three, four and five) corresponding to a gray, green, red and blue interval, respectively.
Combining $\cI_1$ and $\cI_2$, we obtain an \FO successor-graph interpretation in $L$-interval graphs.
\qed

\section{Clique-width of Interval Graphs}
\label{sec:clique}

Every proper hereditary subclass of unit interval graphs has bounded clique-width~\cite{loz08}
though the class of all unit interval graphs has unbounded clique-width~\cite{GR00}.
In particular, the class of $(\{1\},d)$-interval graphs has bounded
clique-width for every $d>0$. Using Gaifman's theorem, it follows that
testing \FO properties of unit interval graphs can be performed in linear time
if the input graph is given by its $\{1\}$-representation
with the left end points of the intervals sorted.
We generalize the result on the clique-width of unit interval graphs for finite sets $L$ of rational numbers,
which proves a special case of our main result for \FO model checking.

\begin{prop}
\label{prop:Ld-boundedcw}
Let $L$ be a finite set of positive rational numbers.
For any $d>0$, the class of $(L,d)$-interval graphs has bounded clique-width.
\end{prop}

\proof
Let $a$ be the largest rational number such that every element of $L$
is an integer multiple of $a$. Without loss of generality, we can assume
that $d$ is not a multiple of $a$ (otherwise, we slightly increase $d$).
We show that the clique-width of any
$(L,d)$-interval graph is at most $K:=\lceil d/a\rceil+1$.

Let $G$ be an $(L,d)$-interval graph with vertices $v_1,\ldots,v_n$ and
fix an $(L,d)$-representation of $G$. Let $b_i$ be the smallest
non-negative real such that $\ell(v_i)-b_i$ is a multiple of $a$.
We may assume that all the numbers $b_i$ are distinct (by perturbing the intervals if needed).
Without loss of generality, we can also assume that $0<b_1<\cdots<b_n<a$.

We will now proceed in several steps. After the $i$-th step,
we will have constructed the subgraph of $G$ induced
by the vertices $v_1,\ldots,v_i$ such that
the label of the vertex $v_i$ is $\lceil\ell(v_i)/a\rceil$.
In the first step, we insert the vertex $v_1$ with label $\lceil\ell(v_1)/a\rceil$.
In the $i$-th step, we insert the vertex $v_i$ with label $K$,
join it by edges to all vertices with labels between $\lceil\ell(v_i)/a\rceil$ and $\lceil r(v_i)/a\rceil$, and
relabel it to $\lceil\ell(v_i)/a\rceil$.
By the choice of $a$ and the assumption that $b_1<\cdots<b_n$,
the vertex $v_i$ is adjacent exactly to its neighbors among $v_1,\ldots,v_{i-1}$. 
\qed

From Proposition~\ref{prop:Ld-boundedcw} and Gaifman's theorem,
one can approach  the \FO model checking problem on $L$-interval graphs
for finite sets $L$ of rationals.
By Gaifman's theorem, every \FO model checking instance can be reduced to
model checking of basic local \FO sentences, i.e.~to \FO model checking on
$L$-interval graphs with bounded radius.
Since $L$-interval graphs with radius $d$ are $(L,(2d+1)\max L)$-interval graphs and so have bounded clique-width,
the latter can be solved in linear time by \cite{cmr00}.
Combining this with the neighborhood covering technique from~\cite{fg01},
which can be adapted to run in linear time in the case of $L$-interval graphs given with their interval representation,
we obtain the following.

\begin{cor}
\label{cor:rationalFO}
Let $L$ be a finite set of positive rational numbers and $\Phi$ an \FO sentence.
There exists a linear time algorithm that decides whether 
an $L$-interval graph $G$ satisfies $\Phi$
if the input graph $G$ is given by its $L$-representation
with the left end points of the intervals sorted.
\end{cor}

However, Proposition~\ref{prop:Ld-boundedcw} is just a fortunate special case, 
since aside of rational lengths one can prove the following.

\begin{prop}
\label{prop:1qcw-unbounded}
For any irrational $q>0$ there is $d$ such that the class of
$\big(\{1,q\},d\big)$-interval graphs has unbounded clique-width.
\end{prop}

\proof
We may assume $q>1$ (otherwise, we rescale and consider the set $\{1,1/q\}$).
We construct a $\big(\{1,q\},d\big)$-interval graph $G$ with arbitrary large clique-width $k$ where $d=q+2$.
Consider a large enough integer $n$; the choice of $n$ depends on $k$ and follows from the construction given.

We construct a sequence $a_1,a_2,\dots,a_n$ of $n$ points from $L^{(n)}\cap [0,d-1)$ as follows:
$a_1=0$, $a_2=1$, and for $i>2$ set
$$a_{i}=\left\{\begin{array}{cl}
               a_{i-1}+1 & \mbox{if $a_{i-1}<d-2$,} \\
	       a_{i-1}-q & \mbox{otherwise.}
	       \end{array}\right.$$
The elements of the sequence defined through the latter case are called {\em $q$-elements}.
Informally, we are folding a sequence of intervals of lengths one and $q$ inside $[0,q+1)$.

Choose $\delta>0$ such that $n\delta$ is smaller than the smallest number in~$L^{(n)}\cap[0,d-1)$.
Let us introduce the following shorthand notation:
if $J$ is an interval and $r$ a real,
then $J+r$ is the interval $J$ shifted by $r$ to the right.
Similarly, if $\II$ is a set of intervals,
then $\II+r$ is the set of the intervals from $\II$ shifted by $r$ to the right.
We define sets of intervals
$\ca U_1:=\{[i\delta,1+i\delta):\> i=0,\dots,n-1\}$ and
$\ca U_q:=\{[i\delta,q+i\delta):\> i=0,\dots,n-1\}$.
We say that intervals $[i\delta,1+i\delta)$ and $[i\delta,q+i\delta)$ are {\em at level~$i$}.

For $i=1,\dots,n$, set $\ca W_i=\ca U_q+a_i$ if $a_i$ is a $q$-element of $P$, and $\ca W_i=\ca U_1+a_i$ otherwise.
Observe that every interval of $\ca W_i$ is a subinterval of $[0,d)$.
Let $G$ be the $L$-interval graph with $n^2$ vertices that
is the intersection graph of the intervals in $\ca W_1\cup\ca W_2\cup\dots\cup\ca W_n$, and
let $W_i$, $i=1,\ldots,n$, be the vertices represented by the intervals from $\ca W_i$.
Finally,
two vertices $x\in W_{i-1}$ and $y\in W_{i}$, $2\leq i\leq n$, are {\em mates}
if they are represented by the same-level intervals.

We claim that the clique-width of $G$ exceeds $k$ if $n$ is sufficiently large.
Suppose that the clique-width of $G$ is at most $k$.
In the construction of $G$ using $k$ labels from the definition of clique-width,
a $k$-labeled subgraph $G_1$ of $G$ with $\frac13n^2\leq |V(G_1)| \leq\frac23n^2$ must have appeared.
However, this implies that vertices of $G_1$ have at most $k$ different neighborhoods in $G\setminus V(G_1)$.
We will show that this is not possible.

Suppose that there exists $i$ such that $|W_{i-1}\cap V(G_1)|-|W_{i}\cap V(G_1)|>k$.
Then there exist $k+1$ vertices in $W_{i-1}\cap V(G_1)$ whose mates are in $W_{i}\sem V(G_1)$ and
these $k+1$ vertices have pairwise distinct neighborhoods in $G\setminus V(G_1)$,
which is impossible.
Similarly, it cannot hold that $|W_{i}\cap V(G_1)|-|W_{i-1}\cap V(G_1)|>k$.

In the rest of the proof, we assume that $||W_{i-1}\cap V(G_1)|-|W_{i}\cap V(G_1)||\le k$ for every $i=2,\ldots,n$.
We say that a set $W_i$ is {\em crossing} if $\emptyset\not=W_i\cap V(G_1)\not=W_i$.
Since we have $\frac13 n^2\le |V(G_1)|\le \frac23 n^2$,
there exist crossing sets $W_{i_0},W_{i_0+1},\dots,$ $W_{i_0+m}$ where $m=\lfloor n/k\rfloor-1$.
If $n$ is large enough,
we can select a $(2k+1)$-element subset $I\subseteq\{i_0,\dots,i_0+m-1\}$
such that neither $a_i$ nor $a_{i+1}$ is a $q$-element for every $i\in I$ (which implies that $a_{i+1}=a_i+1$) and
such that all intervals in $\bigcup_{i\in I}\ca W_i$ share a common point.
Let $i_1,\ldots,i_{2k+1}$ be the elements of $I$ ordered according to the (strictly) increasing values of $a_i$,
i.e.~$a_{i_1}<\cdots<a_{i_{2k+1}}$.

If $j,j'\in\{1,\ldots,2k+1\}$ and $j'>j+1$,
then the neighborhoods of a vertex of $W_{i_j}\cap V(G_1)$ and a vertex of $W_{i_{j'}}\cap V(G_1)$ in $V(G)\sem V(G_1)$ differ.
Indeed, none of the vertices of $W_{i_j}\cap V(G_1)$ is adjacent to any of the vertices in $W_{i_{j+1}+1}\sem V(G_1)$
while each of the vertices of $W_{i_{j'}}\cap V(G_1)$ is adjacent to all the vertices in $W_{i_{j+1}+1}\sem V(G_1)$.
Therefore, the vertices of $G_1$ have at least $k+1$ distinct neighborhoods in $G\setminus V(G_1)$, 
which yields that the clique-width of $G$ is larger than $k$.
\qed

\section{Graph Interpretation in Interval Graphs}
\label{sec:interpret}

This section is devoted to our hardness results concerning model checking for interval graphs.
We first show that Theorem~\ref{thm-L-interval} cannot be generalized to significantly wider classes of interval graphs.
To formulate our results, we need the following definition:
a set $L$ of reals is {\em efficiently dense} in an open set $X$,
if there exists an algorithm that for every non-empty open interval $J\subseteq X$ returns an element of $J\cap L$
in time polynomial in $|J|^{-1}$.

\begin{lem}
\label{lem:1epsFO}
If $L$ is a subset of non-negative reals that is efficiently dense in some non-empty open set,
then there exists an efficient polynomially bounded \FO interpretation of the class of all graphs
in the class of $L$-interval graphs.
\end{lem}
\proof
By scaling, we can assume that $L$ is dense in $[1,1+\varepsilon]$ for some $\varepsilon>0$.
Let $G$ be a graph with $n\ge 2$ vertices (the case $n=1$ is easy to handle separately) and let $v_1,\ldots,v_n$ be its vertices.
We construct an \FO interpretation $\cI=(\nu,\mu)$, which is independent of the choice of $G$, and
an $L$-interval graph $H$ with $3n+5+|E(G)|$ vertices such that $G=\cI(H)$.
We will describe $H$ by giving its $L$-representation.
To simplify our exposition, we assume that $L=[1,1+\varepsilon]$;
it can be routinely verified that the lengths of intervals appearing in the representation of $H$ can be perturbed that
all the length belong to a given dense subset of $[1,1+\varepsilon]$.
Finally, let $\delta=\frac{\varepsilon}{n+1}$.

The vertex set of $H$ will be formed by sets $V_1$, $V_2$ and $V_3$, each containing $n+1$ vertices,
a set $W$ containing $|E(G)|$ vertices, and two special vertices $a$ and $b$.
Let the vertices of $V_i$, $i=1,2,3$, be denoted $t_{i,j}$, $j=0,\ldots,n$, and
the vertices of $W$ be denoted $e_{j,j'}$ for all pairs $1\le j<j'\le n$ such that $v_jv_{j'}\in E(G)$.

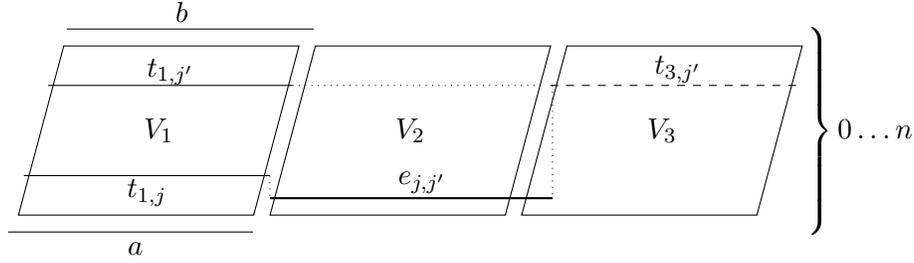
\begin{figure}[tb]
\begin{center}
\begin{tikzpicture}[scale=1.5]
\node at (0,0) [draw,trapezium,trapezium left angle=75,trapezium right angle=-75,minimum height=2.25cm] {$V_1$};
\node at (2.23,0) [draw,trapezium,trapezium left angle=75,trapezium right angle=-75,minimum height=2.25cm] {$V_2$};
\node at (4.46,0) [draw,trapezium,trapezium left angle=75,trapezium right angle=-75,minimum height=2.25cm] {$V_3$};
\draw (-1.335,-0.9) -- (0.835,-0.9);
\node at (-0.205,-1.05) {$a$};
\draw (-0.815,0.9) -- (1.375,0.9);
\node at (0.205,1.05) {$b$};

\draw (-1.195,-0.4) -- (0.985,-0.4);
\node at (-0.105,-0.55) {$t_{1,j}$};
\draw (-0.98,0.4) -- (1.185,0.4);
\node at (0.1025,0.52) {$t_{1,j'}$};
\draw[dashed] (3.47,0.4) -- (5.65,0.4);
\node at (4.6,0.54) {$t_{3,j'}$};
\draw[thick] (0.985,-0.6) -- (3.49,-0.6);
\node at (2.3325,-0.46) {$e_{j,j'}$};
\draw [dotted] (3.49,0.4) -- (3.49,-0.6);
\draw [dotted] (0.985,-0.4) -- (0.985,-0.6);
\draw [dotted] (1.185,0.4) -- (3.4,0.4);
\node at (6.2,0) {$\left.\vbox to 9ex{\vfill}\right\}0\dots n$};
\end{tikzpicture}
\end{center}
\caption{The construction of the interval representation of the graph $H$ in the proof of Lemma~\ref{lem:1epsFO}.}
\label{fig:dense1hardness}
\end{figure}

The vertices of $H$ are represented by the following intervals (also see Figure~\ref{fig:dense1hardness}).
\begin{itemize}
\item The vertex $t_{i,j}$, $i=1,2,3$ and $j=0,1,\dots,n$,
      is represented by the unit interval $\big[i-1+(1+j)\delta,\,i+(1+j)\delta\big)$.
\item The vertex $a$ is represented by the unit interval $\,[0,1)$.
\item The vertex $b$ is represented by the unit interval $\big[(n+2)\delta,\,1+(n+2)\delta\big)$.
\item The $e_{j,j'}\in W$ is represented by the (non-unit) interval $\big[1+j\delta,\,2+(2+j')\delta\big)$.
\end{itemize}
Observe that the vertices $a$ and $t_{1,0}\in V_1$ are {\em twins}, i.e.~they have the same neighbors in $H$, and
that the vertex $b$ is adjacent to every vertex in $V_1\cup V_2\cup\{a\}\cup W$.

Note that the vertices $a$ and $t_{1,0}$ are the only twins in the graph $H$.
In particular, they are the only two vertices that satisfy the following formula:
$$
\prebox{anchor}(x)\equiv\>
 \exists y\, \big(x\not=y\wedge\prebox{edge}(x,y)\wedge \forall z\not=x,y\,
	\prebox{edge}(x,z)\Leftrightarrow\prebox{edge}(y,z)\big)
\,.$$
We will refer to these two vertices as to the {\em anchors}.
Note that the vertices of $V_1$ are at distance one from the anchors,
those of $V_2$ at distance two and
those of $V_3$ at distance three or four.

Let $\prebox{dist}(x,y)=c$ for an integer $c$ be the shorthand for an \FO formula expressing that the distance of two vertices $x$ and $y$ is $c$, and
$\prebox{adist}(x)=c$ for an \FO formula expressing that the distance of $x$ from an anchor is $c$.
The vertices of $G$ are represented by the vertices of $V_1'=\{t_{1,1},\dots,t_{1,n}\}$.
Using this notation, the following formula is true for exactly the vertices of $V'_1$.
$$
\nu(x)\,\equiv\> \neg\prebox{anchor}(x)
 \wedge\prebox{adist}(x)=1\wedge \exists y\,
	(\prebox{adist}(y)=2\wedge\neg\prebox{edge}(x,y))\,.
$$
Note that the last part of the formula makes $\nu(x)$ false for $x=b$.

In what follows, we refer to the pairs of vertices $t_{1,j}$ and $t_{3,j}$ as \emph{mates}.
The following formula is true if and only if $x'\in V_3$ is the mate of $x\in V_1'$:
\begin{align*}
\prebox{mates}(x,x')\,\equiv~&
 \nu(x)\wedge(\prebox{adist}(x')=3\lor\prebox{adist}(x')=4)\wedge
\\
	&\exists!y\, \big( \prebox{adist}(y)=2\wedge
		\neg\prebox{edge}(x,y)\wedge\neg\prebox{edge}(x',y)
	\big).
\end{align*}
Suppose that $x=t_{1,j}$ and $x'=t_{3,j'}$.
If $j'<j$, then there exists no vertex $y$ as in the formula and,
if $j'>j$, there exists at least two such $y$'s, in particular, $t_{2,j},\dots,t_{2,j'}$.

The vertices of $V_1'$ can be linearly ordered according to their left end points.
This linear order is actually reflected by dominating one vertex of another.
Formally, a vertex $x$ {\em dominates} a vertex $y$ if $y$ and all its neighbors are also neighbors of~$x$.
Observe that $x\in V'_1$ dominates $y\in V_1'$ if and only if the left end point of $y$ precedes the left end points of $x$.
The following \FO formula expresses that a vertex $x$ dominates a vertex $y$.
$$\prebox{domin}(x,y)\,\equiv\> x\not=y\wedge\prebox{edge}(x,y)\wedge \forall z\,(\prebox{edge}(y,z)\to\prebox{edge}(x,z))\,.$$
Using this formula, we can define the formula $\mu$.
\begin{align*}
\mu(x,y)\,\equiv&\> \mu'(x,y)\vee\mu'(y,x) \,,
\mbox{~~where}\\
\mu'(x,y)\,\equiv&\, \prebox{domin}(y,x)\wedge 
	\exists\, y',z\;\big[ \prebox{mates}(y,y')\wedge
\\
&\qquad\prebox{edge}(x,z)\wedge\forall
		t\,(\prebox{domin}(x,t)\to\neg\prebox{edge}(t,z)) \wedge
\\
&\qquad\prebox{edge}(y',z)\wedge\forall
		t\,(\prebox{domin}(y',t)\to\neg\prebox{edge}(t,z)) \,
\big].
\end{align*}
Note that $\mu'(x,y)$ for $x=t_{1,j}$ and $y=t_{1,j'}$ is true if and only if $j<j'$ and the set $W$ contains the vertex $e_{j,j'}$.
Indeed, $z=e_{j,j'}$ is the only possible choice of a vertex satisfying the existential quantification.
\qed

Since the parameterized \FO model checking problem is AW[*]-complete
for general graphs, we can immediately conclude the following.
\begin{cor}
\label{cor-FO}
If $L$ is a subset of non-negative reals that is efficiently dense in some non-empty open set,
then \FO model checking is AW[*]-complete on $L$-interval graphs when
parameterized by the formula size.
\end{cor}

We now turn our attention to interpretations in stronger logics.
We start by showing that the class of all graphs
has an \FO interpretation in the class of unit interval graphs with a successor relation.
We actually prove a stronger statement that there exists an interpretation of the class of all directed graphs.

\begin{lem}
\label{lm-succ-lower}
There exists a polynomially bounded \FO interpretation of the class of all directed graphs
in the class of unit interval graphs with a successor relation.
\end{lem}

\proof
Fix a directed graph $G$.
Let $n$ and $m$ be the number of vertices and edges of $G$, respectively.
Further, let $v_1,\ldots,v_n$ be the vertices of $G$,
let $d^+_i$ and $d^-_i$ be the out-degree and in-degree of a vertex $v_i$ and
let $e_{i,1},\ldots,e_{i,d^+_i}$ be the edges leaving $v_i$.
We will simultaneously describe the \FO interpretation $\cI=(\nu,\mu)$ and an unit interval graph $H$ such that $G=\cI(H)$.

For each vertex $v_i$ of $G$,
the graph $H$ contains the following $2+d^+_i+d^-_i$ vertices:
$u_i$, $u'_i$ and $u_{i,e}$ for each edge $e$ leaving or entering $v_i$ in $G$.
The graph $H$ consists of $n$ cliques, the $i$-th clique formed by the $2+d^+_i+d^-_i$ vertices corresponding to $v_i$.
Clearly, $H$ is a unit interval graph.

We now define a successor relation on the vertices of $H$.
To make the definition of the successor relation less technical,
we abuse the notation by writing $u_{i,e_{i,0}}$ for $u'_i$ (note that there is no edge denoted by $e_{i,0}$ in $G$).
The successor relation will contain the following pairs of vertices of $H$:
\begin{itemize}
\item $(u_i,u'_i)=(u_i,u_{i,e_{i,0}})$ for every $i=1,\ldots,n$,
\item $(u_{i,e_{i,j-1}},u_{i',e_{i,j}})$ and $(u_{i',e_{i,j}},u_{i,e_{i,j}})$ for every edge $e_{i,j}$, $i=1,\ldots,n$ and $j=1,\ldots,d^+_i$,
      where $u_{i'}$ is the head of $e_{i,j}$, and
\item $(u_{i,e_{i,d^+_i}},u_{i+1})$ for every $i=1,\ldots,n-1$.
\end{itemize}
Note that the only pairs of adjacent vertices included in the successor relation are those described in the first item.
The following two \FO formulas can be used to form the interpretation.
\begin{align*}
\nu(x)   \> \equiv \> & \exists t \; \prebox{succ}(x,t)\wedge \prebox{edge}(x,t) \\
\mu(x,y) \> \equiv \> & \exists t,t',t'' \; \prebox{succ}(t,t')\wedge \prebox{succ}(t',t'') \wedge \\
                      & \prebox{edge}(x,t) \wedge \prebox{edge}(y,t') \wedge \prebox{edge}(x,t'').
\end{align*}
It is straightforward to check that $G=\cI(H)$.
\qed

Lemma~\ref{lm-succ-lower} yields the following.

\begin{cor}
\label{cor-succFO}
\FO model checking is AW[*]-complete on unit interval graphs with a successor relation
when parameterized by the formula size.
\end{cor}

We now turn our attention to more general \MSO properties.
There exist two commonly used \MSO frameworks for graphs:
the \MSOi language where quantifying over vertices and vertex sets only is allowed, and
\MSOii where it is allowed to quantify over edges and edge sets in addition.
Our negative result holds for the weaker variant \MSOi (and so also holds for \MSOii).

\begin{lem}
\label{lem:1MSO1}
There is a polynomially bounded \MSOi interpretation of the class of all graphs in the class of unit interval graphs.
\end{lem}

\begin{figure}[tb]
\begin{center}
\begin{tikzpicture}[scale=1.6]
\node at (0.7,0) [draw,rectangle, xslant=0.3, minimum height=2cm, minimum width= 1.5cm] {$W_0$};
\node at (1.8,0) [draw,rectangle, xslant=0.3, minimum height=2cm, minimum width= 1.5cm] {};
\node at (2.9,0) [draw,rectangle, xslant=0.3, minimum height=2cm, minimum width= 1.5cm] {};
\node at (4.0,0) [draw,rectangle, xslant=0.3, minimum height=2cm, minimum width= 1.5cm] {$W_1$};

\draw (-0.55,-0.8) -- (0.45,-0.8);
\draw (-1.05,-1) -- (-0.05,-1);
\draw (-1.05,-1.06) -- (-0.05,-1.06);
\draw (-1.05,-0.94) -- (-0.05,-0.94);
\node at (-1.2, -1) {$A$};
\draw (1.675,-1) -- (2.675,-1);
\draw (1.675,-1.06) -- (2.675,-1.06);
\draw (1.675,-0.94) -- (2.675,-0.94);
\node at (1.48, -1) {$A_1$};

\draw (0.07,-0.45) -- (1.07,-0.45);
\draw[dotted] (1.17,-0.45) -- (3.27,-0.45);
\node at (0.57,-0.3) {\small level $i$};
\draw (0.33,0.45) -- (1.33,0.45);
\draw[dotted] (1.43,0.45) -- (3.53,0.45);
\node at (0.76,0.33) {\small level $j$};

\draw[dashed] (3.37,-0.45) -- (4.37,-0.45);
\draw[dashed] (3.63,0.45) -- (4.63,0.45);
\node at (3.87,-0.3) {\small level $i$};
\node at (4.06,0.33) {\small level $j$};
\draw [dotted] (3.37,-0.45) -- (3.37,-0.8);
\draw [dotted] (4.37,-0.45) -- (4.37,-0.8);
\draw (3.37,-0.8) -- (4.37,-0.8);
\draw [dotted] (3.63,0.45) -- (3.63,-0.9);
\draw [dotted] (4.63,0.45) -- (4.63,-0.9);
\draw (3.63,-0.9) -- (4.63,-0.9);

\tikzstyle{every node}=[draw, shape=circle, minimum size=2pt,inner sep=0pt, fill=black]
\node (a) at (5.5,0) {};
\node (a) at (5.3,0) {};
\node (a) at (5.1,0) {};

\end{tikzpicture}
\end{center}
\caption{The interval representation of the graph $H$ with a part representing an edge $v_iv_j$ of the graph $G$.}
\label{fig:unit1hardness}
\end{figure}
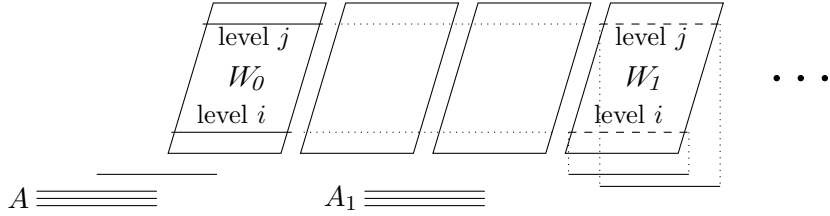

\proof
We describe the \MSOi interpretation $\cI=(\nu,\mu)$.
Fix an $n$-vertex $G$ with $n\ge 4$ (the cases with $n=1,2,3$ can be handled separately in a straightforward way).
Let $v_1,\ldots,v_n$ be the vertices of $G$ and $e_1,\ldots,e_m$ its edges.
We will construct a unit interval graph $H$ such that $G=\cI(H)$.
The graph $H$ will be described by giving its interval representation and
its construction is illustrated in Figure~\ref{fig:unit1hardness}.

Choose $\delta>0$ such that $\delta n<\frac12$ and
$\ca U=\big\{\, [i\delta,1+i\delta):\, i=0,1,\dots,n-1\big\}$.
Recall that $J+x$ where $J$ is an interval and $x$ is a real
is the interval $J$ shifted by $x$ to the right.
The graph $H$ contains $n(3m+1)$ vertices corresponding to the intervals from the sets $\ca U+k$ for $k=1,\ldots,3m+1$;
the vertices corresponding to the intervals $[i\delta,1+i\delta)$ and $[i\delta,1+i\delta)+k$ are said to be at the {\em level $i$}.
Let $W_\ell$, $\ell=0,\ldots,m$, be the set of the $n$ vertices represented by the intervals from $\ca U+(3\ell+1)$.

The graph $H$ further contains three vertices represented by the interval $[0,1)$ each and
$m$ triples of vertices represented by the intervals $[0,1)+(3i-1/2)$, $i=1,\ldots,m$.
The vertices in these $m+1$ triples will be referred to as {\em anchors} and
they will be the only vertices of $H$ that have two twins.
Also insert a vertex represented by the interval $[1/2,3/2)$.
The three vertices represented by the interval $[0,1)$ are the only anchors of degree four.

If the edge $e_k$ joins vertices $v_i$ and $v_j$,
$H$ contains a pair of vertices represented by the intervals
\mbox{$[i\delta,1+i\delta)+(3k+1)$} and $[j\delta,1+j\delta)+(3k+1)$.
The vertices included in this step are the only vertices of $H$ that have unique twins.
This finishes the construction of $H$.

We now give the \MSOi formulas $\nu$ and $\mu$.
Let $\prebox{twin}(x,y)$ be the \FO formula expressing that $x$ and $y$ are twins.
Using this formula, we can identify the anchors and vertices not adjacent to any of the anchors.
\begin{align*}
\prebox{anchor}(x)\equiv\>&
 \exists y,z\, \big(z\not=y\wedge\prebox{edge}(z,y)\wedge 
	\prebox{twin}(x,y)\wedge\prebox{twin}(x,z)\big)
\,,\\
\prebox{noanch}(x)\equiv\>&
 \forall z\, \big(\prebox{anchor}(z)\Rightarrow \neg\prebox{edge}(x,z)\big).
\end{align*}
Note that the only vertices $x$ that satisfy $\prebox{noanch}(x)$ are 
the vertices in the sets $W_1,\ldots,W_m$ and the $2m$ twins corresponding to the edges of $G$.
The vertices of $G$ will be modeled by the vertices of $W_0$,
which are precisely the vertices that are not adjacent to any anchor and
that are at distance two from the three anchors of degree four.
In particular, the formula $\nu$ can be chosen to be the following \FO formula.
$$
\nu(x)\,\equiv\> \prebox{noanch}(x)\wedge\, \exists t
	\big( \prebox{anchor}(t)\wedge \prebox{deg}(t)=4\wedge 
		\prebox{dist}(t,x)=2 \big)
\,.$$

Two vertices $x$ and $x'$ are {\em mates}
if there exists integers $p$, $1\le p\le m$, and $i$, $0\le i\le n-1$, such that
one of them is represented by the interval $[i\delta,1+i\delta)+(3p-2)$ and
the other is represented by the interval $[i\delta,1+i\delta)+(3p+1)$.
In particular, if $x\in W_{p-1}$ and $x'\in W_p$ and the vertices $x$ and $x'$ are represented by intervals at the same level,
then $x$ and $x'$ are mates.
It is easy to verify that two vertices $x$ and $x'$ are mates iff they satisfy the following \FO formula.
\begin{align*}
\prebox{mates}(x,x')\,\equiv~&
 \prebox{noanch}(x)\wedge\prebox{noanch}(x')\wedge\prebox{dist}(x,x')=4\wedge
	\exists t\,\big[\prebox{anchor}(t)\wedge 
\\
	&\exists!y\,\exists!z
	 \big( \neg\prebox{edge}(y,z)\wedge\prebox{edge}(y,t)
		\wedge\prebox{edge}(z,t) \wedge \prebox{dist}(x,y)=2\wedge
\\
	&\prebox{dist}(x',y)>2\wedge
	   \prebox{dist}(x',z)=2\wedge\prebox{dist}(x,z)>2
	\big)\big]\,.
\end{align*}
The transitive closure of the binary relation given by $\prebox{mates}$ can be described
by the following \MSO formula $\prebox{mates}^*(x,y)$.
\begin{align*}
\prebox{mates}{}^*(x,x')\,\equiv~&
   x=x'\vee \exists U\; \left[x\in U\wedge x'\in U\wedge \exists! t\in U\;\prebox{mates}(x,t)\wedge\right. \\
   &\exists! t\in U\;\prebox{mates}(x',t)\wedge \forall y\in U (x\not=y\wedge x'\not=y)\Rightarrow \\
   & \left.(\exists t\in U\; \exists! t'\in U\; t\not=t'\wedge \prebox{mates}(y,t)\wedge\prebox{mates} (y,t')) \right]\,.
\end{align*}
Note that this is the only place in the proof where we need the expressive power of \MSO.

The formula $\mu$ can now be chosen as follows.
\begin{align*}
\mu(x,y)\,\equiv~&\> x\not=y\wedge
	\exists x',x'',y',y'' \big( \prebox{edge}(x',y')\wedge
\\
	&\, \prebox{mates}{}^*(x,x')\wedge\prebox{mates}{}^*(y,y')\wedge
		\prebox{twin}(x',x'')\wedge\prebox{twin}(y',y'') \big)
\,.\end{align*}
Indeed, if $x$ and $y$ belong to $W_0$,
then $\mu(x,y)$ is true only if there exist adjacent vertices $x'$ and $y'$ at the same level as $x$ and $y$, respectively, and
both $x'$ and $y'$ have twins.
However, this happens only if the counterparts of $x$ and $y$ in $G$ are joined by an edge.
\qed

Hence we obtain the following.
\begin{cor}
\label{cor-MSO}
\MSOi model checking is para-\textup{NP}-hard on unit interval graphs.
\end{cor}
Note that the aforementioned result of Lozin~\cite{loz08} states that
every proper hereditary subclass of unit interval graphs has bounded clique-width,
and hence \MSOi model checking on this class can be carried out in linear time~\cite{cmr00}.

\bibliographystyle{alpha}
\bibliography{INT-FO-lmcs}

\end{document}